\documentclass[a4paper,fleqn]{cas-dc}

\usepackage[numbers]{natbib}
\usepackage[T1]{fontenc}
\usepackage{listings}
\usepackage{xcolor}
\usepackage{booktabs}
\usepackage{tikz}
\usetikzlibrary{arrows.meta, positioning, shapes.geometric, fit}
\usepackage{amsmath,amssymb}
\usepackage{stfloats}
\usepackage{academicons}
\usepackage{fontawesome5}
\definecolor{orcidlogocol}{rgb}{0.65, 0.807, 0.223}
\newcommand{\orcid}[1]{$\,$\href{https://orcid.org/#1}{\textcolor{orcidlogocol}{\faOrcid}}}

\makeatletter
\renewcommand\section{\@startsection{section}{1}{\z@}%
    {15pt \@plus 3\p@ \@minus 3\p@}%
    {6\p@}%
    {\sectionfont\raggedright\hst[13pt]}}
\renewcommand\subsection{\@startsection{subsection}{2}{\z@}%
    {10pt \@plus 3\p@ \@minus 2\p@}%
    {4\p@}%
    {\ssectionfont\raggedright}}
\renewcommand\subsubsection{\@startsection{subsubsection}{3}{\z@}%
    {10pt \@plus 1\p@ \@minus .3\p@}%
    {4\p@}%
    {\sssectionfont\raggedright}}
\makeatother

\definecolor{codebg}{gray}{0.95}
\definecolor{codegreen}{rgb}{0.0,0.4,0.0}
\definecolor{codegray}{rgb}{0.5,0.5,0.5}
\definecolor{codepurple}{rgb}{0.5,0.0,0.35}
\lstdefinestyle{nanoCMB}{
  language=Python,
  backgroundcolor=\color{codebg},
  basicstyle=\ttfamily\footnotesize,
  keywordstyle=\color{codepurple}\bfseries,
  commentstyle=\color{codegreen},
  stringstyle=\color{blue!60!black},
  numberstyle=\tiny\color{codegray},
  numbers=left,
  numbersep=6pt,
  frame=single,
  framerule=0.4pt,
  rulecolor=\color{gray!50},
  breaklines=true,
  breakatwhitespace=false,
  showstringspaces=false,
  columns=flexible,
  xleftmargin=2em,
  aboveskip=0.8em,
  belowskip=0.8em,
  morekeywords={as, True, False, None},
  literate={τ}{{\ensuremath{\tau}}}1 {η}{{\ensuremath{\eta}}}1 {π}{{\ensuremath{\pi}}}1
           {²}{{\textsuperscript{2}}}1 {⁻}{{\textsuperscript{-}}}1
           {¹}{{\textsuperscript{1}}}1 {⁴}{{\textsuperscript{4}}}1
           {₀}{{\textsubscript{0}}}1 {₁}{{\textsubscript{1}}}1
}

\begin{document}
\let\WriteBookmarks\relax
\def\floatpagepagefraction{1}
\def\textpagefraction{.001}

\shorttitle{nanoCMB: A minimal CMB power spectrum calculator}
\shortauthors{A.\ Moss}

\title [mode = title]{nanoCMB: A minimal CMB power spectrum calculator in Python}

\author[1]{Adam Moss\orcid{0000-0002-7245-7670}}
\cormark[1]
\ead{adam.moss@nottingham.ac.uk}

\credit{Conceptualization, Methodology, Software, Writing}

\affiliation[1]{organization={School of Physics and Astronomy, University of Nottingham},
            city={Nottingham},
            postcode={NG7 2RD},
            country={United Kingdom}}

\cortext[1]{Corresponding author}

\begin{abstract}
We present \texttt{nanoCMB}, a minimal but accurate calculator for the unlensed
CMB temperature and polarisation angular power spectra ($C_\ell^{TT}$,
$C_\ell^{EE}$, $C_\ell^{TE}$) of flat $\Lambda$CDM cosmologies.  Written in
$\sim$1400 lines of readable Python, the code implements the full
line-of-sight integration method: RECFAST recombination, coupled
Einstein--Boltzmann perturbation equations in synchronous gauge with a
tight-coupling approximation, precomputed spherical Bessel function tables,
and optimally constructed non-uniform grids in wavenumber and conformal time.
Despite its brevity, \texttt{nanoCMB} achieves sub-percent agreement with CAMB
across the multipole range $2 \le \ell \le 2500$, running
in $\sim$10 seconds on a modern laptop.  The entire calculation lives in a single, easily modifiable Python script,
designed as a pedagogical bridge between textbook treatments and
research-level Boltzmann solvers, with every approximation and numerical
choice made explicit.  We describe the physics, equations, and computational methods in
detail, with code snippets illustrating each stage of the calculation.
The code is publicly available at \url{https://github.com/adammoss/nanoCMB}.
\end{abstract}


\maketitle

\section{Introduction}
\label{sec:intro}

The cosmic microwave background (CMB) angular power spectrum is a cornerstone
of precision cosmology.  Measurements by COBE, WMAP, and Planck have
determined the parameters of the $\Lambda$CDM model to percent-level accuracy
or better~\cite{Planck:2018vyg}, and upcoming experiments will push these
constraints further.  Computing the power spectrum from a set of cosmological
parameters requires solving the coupled Einstein--Boltzmann equations through
recombination and integrating the resulting source functions along the line
of sight.  This calculation was first implemented in
COSMICS~\cite{Bertschinger:1995er} and CMBFAST~\cite{Seljak:1996is}, and
is now handled by the research-level codes
CAMB~\cite{Lewis:1999bs} and CLASS~\cite{Blas:2011rf}, which underpin
essentially all modern CMB parameter estimation~\cite{Planck:2018vyg}.

These codes are impressive pieces of software, but they are also large ---
tens of thousands of lines of Fortran or C, with many layers of optimisation
and historical conventions.  For someone new to the field, they are not easy to learn from. Textbooks
like Dodelson \& Schmidt~\cite{Dodelson:2020bqr} explain the physics well,
but there is a real gap between the equations on the page and a working code:
choosing a gauge, setting up grids, implementing tight coupling, handling
stiff ODEs, constructing source functions.  These practical details matter
enormously for accuracy but are rarely discussed.

\texttt{nanoCMB} tries to bridge this gap.  It computes the unlensed $TT$,
$EE$, and $TE$ power spectra for flat $\Lambda$CDM cosmologies in
$\sim$1400 lines of Python, with sub-percent accuracy relative to CAMB.
The code follows the physics directly: background $\to$ recombination $\to$
perturbations $\to$ line-of-sight integration $\to$ power spectra, all in a
single file with no class hierarchies or configuration layers.
Every approximation is spelled out and every numerical choice is visible.

Because everything is in one readable file, it is easy to modify.
Swapping in a different recombination model, changing the tight-coupling
scheme, or adding a new source term means editing a few functions rather
than navigating a large package.  This makes \texttt{nanoCMB} useful both
for teaching and for quickly testing new ideas.

The closest predecessor is the pedagogical code of
Callin~\cite{Callin:2006qx}, a Fortran implementation of the temperature
anisotropy calculation.  \texttt{nanoCMB} adds polarisation, full RECFAST
recombination with helium, and optimised non-uniform grids, while staying
in Python.

We restrict ourselves to unlensed scalar modes in flat cosmologies with
massless neutrinos and $w = -1$.  There is no lensing, no tensors, no
massive neutrinos, no curvature, and no isocurvature modes.  These
limitations keep the code short; the modular structure should make
extensions reasonably straightforward.

The remainder of this paper follows the structure of the code itself.
Section~\ref{sec:background} describes the background cosmology.
Section~\ref{sec:recombination} covers recombination and the ionisation
history.  Section~\ref{sec:perturbations} presents the Boltzmann hierarchy
and its numerical solution.  Section~\ref{sec:los} describes the
line-of-sight integration method.  Section~\ref{sec:cls} covers the power
spectrum assembly.  Section~\ref{sec:grids} discusses the optimal non-uniform
grid construction.  Section~\ref{sec:validation} validates the final
power spectra against CAMB.  Section~\ref{sec:conclusions} concludes.
The code is publicly available at
\url{https://github.com/adammoss/nanoCMB}.

\section{Background cosmology}
\label{sec:background}

\subsection{The Friedmann equation}
\label{sec:friedmann}

We work in conformal time $\tau$ with the flat FRW metric
$ds^2 = a^2(\tau)(-d\tau^2 + d\mathbf{x}^2)$, in units where $c = 1$ and
distances are measured in Mpc.  The Hubble parameter is defined as usual
with respect to cosmic time $t$,
\begin{equation}
\label{eq:friedmann}
H \equiv \frac{1}{a}\frac{da}{dt} = \frac{\dot{a}}{a^2}\,,\qquad
H^2 = \frac{8\pi G}{3}\,\rho_{\rm tot}\,,
\end{equation}
where overdots denote derivatives with respect to conformal time
($\dot{a} \equiv da/d\tau = a^2 H$), and $\rho_{\rm tot}$
is the total energy density.

Following CAMB convention, we define the ``grhoa2'' combination
\begin{equation}
\label{eq:grhoa2}
\mathcal{G}(a) \equiv 8\pi G\,\rho_{\rm tot}\,a^4
= \bar{\rho}_\gamma + \bar{\rho}_\nu + (\bar{\rho}_c + \bar{\rho}_b)\,a + \bar{\rho}_\Lambda\,a^4\,,
\end{equation}
where $\bar{\rho}_i$ are the present-day density values.
The Hubble parameter and conformal time derivative are then
\begin{equation}
\label{eq:hubble}
H(a) = \frac{\sqrt{\mathcal{G}(a)/3}}{a^2}\,,\qquad
\frac{d\tau}{da} = \sqrt{\frac{3}{\mathcal{G}(a)}}\,.
\end{equation}

In the code, these are implemented concisely:

\begin{lstlisting}[style=nanoCMB, caption={Friedmann equation building blocks.}]
def grhoa2(a, bg):
    """Total 8piG*rho*a^4."""
    return (bg["grhog"] + bg["grhornomass"]
            + (bg["grhoc"] + bg["grhob"]) * a
            + bg["grhov"] * a**4)

def dtauda(a, bg):
    """d(tau)/da = 1/(a^2 H) = sqrt(3/grhoa2) in Mpc."""
    return np.sqrt(3.0 / grhoa2(a, bg))
\end{lstlisting}

The individual density parameters are computed from the input cosmological
parameters (table~\ref{tab:params}).  The photon density is derived from the
CMB temperature $T_{\rm CMB}$ via $\rho_\gamma = (4\sigma_{\rm SB}/c)\,T_{\rm CMB}^4$,
and the massless neutrino density follows from the standard relation
$\rho_\nu = (7/8)(4/11)^{4/3} N_{\rm eff}\,\rho_\gamma$.  The cosmological
constant is fixed by the flatness condition $\Omega_\Lambda = 1 - \Omega_m - \Omega_r$.

\subsection{Conformal time and the sound horizon}
\label{sec:conformal}

The conformal time is obtained by direct numerical integration:
\begin{equation}
\label{eq:conformal_time}
\tau(a) = \int_0^a \frac{da'}{a'^2\,H(a')} = \int_0^a \sqrt{\frac{3}{\mathcal{G}(a')}}\,da'\,.
\end{equation}
The age of the universe in conformal time, $\tau_0 = \tau(a=1)$, and the
conformal time at matter--radiation equality, $\tau_{\rm eq}$, are key derived
quantities.

The comoving sound horizon in the photon--baryon fluid is
\begin{equation}
\label{eq:sound_horizon}
r_s(a) = \int_0^a \frac{c_s\,da'}{a'^2\,H(a')}\,,\qquad
c_s^2 = \frac{1}{3(1+R)}\,,
\end{equation}
where $R = 3\rho_b/(4\rho_\gamma) = (3\bar{\rho}_b\,a)/(4\bar{\rho}_\gamma)$
is the baryon loading.  The sound horizon at recombination, $r_s(a_*)$,
sets the angular scale of the acoustic peaks via $\theta_* = r_s / D_A(z_*)$.

\begin{lstlisting}[style=nanoCMB, caption={Sound horizon integrand.}]
def sound_horizon(a, bg):
    """Comoving sound horizon r_s(a) = int_0^a c_s da'/(a'^2 H)."""
    def integrand(ap):
        R = 0.75 * bg["grhob"] * ap / bg["grhog"]
        return 1.0 / np.sqrt(grhoa2(ap, bg) * (1.0 + R))
    return integrate.quad(integrand, 0, a)[0]
\end{lstlisting}

\begin{figure*}[!t]
\centering
\includegraphics[width=\textwidth]{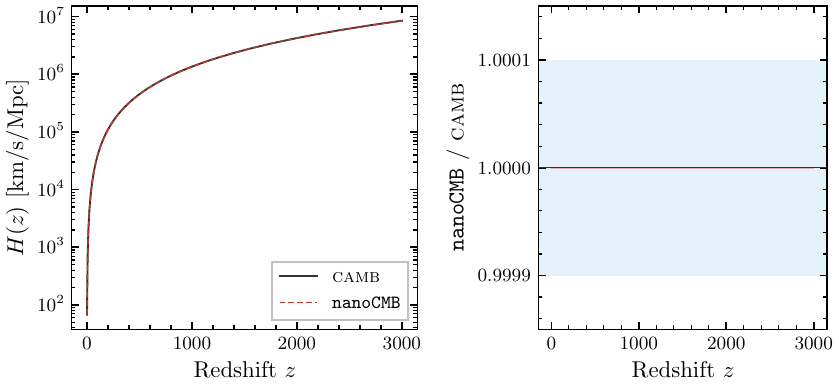}
\caption{\label{fig:background}
Hubble parameter comparison.  {\em Left:} $H(z)$ from \texttt{nanoCMB}
(red dashed) and CAMB (black solid).  {\em Right:} ratio, showing agreement
at the $10^{-4}$ level.}
\end{figure*}

Figure~\ref{fig:background} compares the Hubble parameter $H(z)$ from
\texttt{nanoCMB} to CAMB, demonstrating agreement at the $10^{-4}$ level
across all redshifts.

\subsection{Unit conventions}
\label{sec:units}

All distances are in Mpc, with $c = 1$ so that times are also in Mpc.
The Hubble parameter $H$ has units of Mpc$^{-1}$, and wavenumbers $k$ have
units of Mpc$^{-1}$.  Energy densities are stored in the CAMB
convention as $8\pi G\rho$ (units of Mpc$^{-2}$), so that the Friedmann
equation involves no explicit factors of $G$.  The conversion
from the conventional $H_0 = 100\,h\;\text{km/s/Mpc}$ is
$H_0 = 100\,h / c_{\rm km/s}$ in code units.

\begin{table}[tbp]
\centering
\begin{tabular}{llc}
\toprule
Parameter & Symbol & Value \\
\midrule
Baryon density & $\omega_b \equiv \Omega_b h^2$ & 0.02237 \\
CDM density & $\omega_c \equiv \Omega_c h^2$ & 0.1200 \\
Hubble parameter & $h$ & 0.6736 \\
Scalar spectral index & $n_s$ & 0.9649 \\
Scalar amplitude & $A_s$ & $2.1 \times 10^{-9}$ \\
Reionisation optical depth & $\tau_{\rm reion}$ & 0.0544 \\
Effective neutrino number & $N_{\rm eff}$ & 3.044 \\
CMB temperature & $T_{\rm CMB}$ & 2.7255\,K \\
Helium fraction & $Y_{\rm He}$ & 0.245 \\
Pivot scale & $k_0$ & 0.05\,Mpc$^{-1}$ \\
\bottomrule
\end{tabular}
\caption{\label{tab:params}
Cosmological parameters used throughout this work, corresponding to the
Planck 2018 best-fit $\Lambda$CDM model~\cite{Planck:2018vyg}.}
\end{table}

\section{Recombination}
\label{sec:recombination}

\subsection{Physics of hydrogen recombination}
\label{sec:hydrogen}

As the universe cools below $T \sim 4000\,$K, the ionisation fraction
$x_e = n_e / n_H$ drops from unity as electrons and protons combine to form
neutral hydrogen.  The na\"ive expectation from the Saha equation is that
recombination occurs at $T \sim 3700\,$K, but the Saha equation assumes
thermal equilibrium, which fails because Lyman-$\alpha$ photons produced
during recombination are immediately reabsorbed, creating a bottleneck.

The effective three-level atom model of Peebles~\cite{Peebles:1968ja} (see also
Zeldovich, Kurt \& Sunyaev~\cite{Zeldovich1968}) resolves this by tracking
only the ground state and the $n=2$ shell.  Recombinations directly to the
ground state produce photons that immediately ionise another atom, so the
{\em effective} recombination rate is the case-B rate $\alpha_B$ (to excited
states only).  The two escape routes from $n=2$ are Lyman-$\alpha$ escape
(net rate suppressed by the Sobolev optical depth) and two-photon decay
($2s \to 1s$, rate $\Lambda_{2s \to 1s} = 8.22\,$s$^{-1}$).

The Peebles equation for hydrogen is
\begin{multline}
\label{eq:peebles}
\frac{dx_H}{dz} = \frac{C_r}{H(z)(1+z)}\Big[
x_e\,x_H\,n_H\,\alpha_B \\
- \beta_B\,(1-x_H)\,e^{-E_{21}/k_BT_m}
\Big]\,,
\end{multline}
where $x_H = n_{H^+}/n_H$ is the hydrogen ionisation fraction,
$\alpha_B$ is the case-B recombination coefficient~\cite{Pequignot1991},
$\beta_B = \alpha_B\,(2\pi m_e k_B T_m / h^2)^{3/2}\,e^{-B_2/k_BT_m}$
is the photoionisation rate from $n=2$, $E_{21}$ is the Lyman-$\alpha$ energy,
and the Peebles $C$-factor is
\begin{equation}
\label{eq:Cr}
C_r = \frac{1 + K\,\Lambda_{2s1s}\,n_{1s}}{1 + K\,(\Lambda_{2s1s} + \beta_B)\,n_{1s}}\,,
\end{equation}
with $K = \lambda_\alpha^3 / (8\pi H)$ encoding the Sobolev optical depth
of the Lyman-$\alpha$ line, and $n_{1s} = (1-x_H)\,n_H$ the ground-state
number density.

\subsection{RECFAST implementation}
\label{sec:recfast}

\texttt{nanoCMB} implements the full RECFAST algorithm~\cite{Seager:1999bc,
Seager:1999bc,Seager:1999km,Wong:2007ym}, which extends the Peebles equation to include helium
recombination (both He\,{\sc iii}$\to$He\,{\sc ii} and He\,{\sc ii}$\to$He\,{\sc i},
including singlet and triplet channels with hydrogen continuum opacity
corrections) and tracks the matter temperature $T_m$ separately from the
radiation temperature $T_r = T_{\rm CMB}(1+z)$.

The calculation proceeds in phases, from high to low redshift:
\begin{enumerate}
\item $z > 8000$: fully ionised ($x_e = 1 + 2f_{\rm He}$, helium doubly ionised).
\item $5000 < z < 8000$: He$^{++} \to$ He$^+$ via Saha equilibrium.
\item $3500 < z < 5000$: He singly ionised, H fully ionised ($x_e = 1 + f_{\rm He}$).
\item $z < 3500$: He$^+ \to$ He$^0$ in Saha equilibrium until
$x_{\rm He} < 0.99$, then a full three-variable ODE system
$(x_H, x_{\rm He}, T_m)$ solved with implicit Radau IIA, which
handles the hydrogen transition automatically.
\end{enumerate}

The hydrogen piece of the ODE right-hand side illustrates the Peebles
equation in code:

\begin{lstlisting}[style=nanoCMB, caption={Hydrogen Peebles equation from \texttt{recfast\_rhs()}.}]
# Case-B recombination coefficient (Pequignot et al. 1991 fit)
t4 = T_mat / 1e4
Rdown = 1e-19 * 4.309 * t4**(-0.6166) / (1 + 0.6703 * t4**0.5300)
Rup = Rdown * (CR * T_mat)**1.5 * np.exp(-CDB / T_mat)

# Sobolev escape: K = lambda_alpha^3/(8pi*H) with Hswitch correction
K = CK / Hz * (1.0
    + AGauss1 * np.exp(-((np.log(1+z) - zGauss1) / wGauss1)**2)
    + AGauss2 * np.exp(-((np.log(1+z) - zGauss2) / wGauss2)**2))
n_1s = n_H * max(1 - x_H, 1e-30)

f1 = ((x * x_H * n_H * Rdown
       - Rup * (1 - x_H) * np.exp(-CL / T_mat))
      * (1 + K * Lambda_2s1s * n_1s)
      / (Hz * (1+z) * (1.0/fu + K * Lambda_2s1s * n_1s / fu
                        + K * Rup * n_1s)))
\end{lstlisting}

The RECFAST fudge factor $f_u = 1.125$ and the Hswitch double-Gaussian
correction to $K$ (parameterised by amplitudes $A_{1,2}$, centres
$z_{1,2}$, and widths $w_{1,2}$) follow the CAMB defaults~\cite{Wong:2007ym}.

The helium singlet channel follows an analogous Peebles equation with
the He\,{\sc i} case-B coefficient of Hummer \& Storey~\cite{HummerStorey1998},
Sobolev escape via the $2^1P_1 \to 1^1S_0$ line, two-photon decay from
$2^1S_0$ (rate $\Lambda_{\rm He} = 51.3\,$s$^{-1}$), and hydrogen
continuum opacity corrections to the effective escape
probability~\cite{Seager:1999bc,Wong:2007ym}.  The triplet channel
(He\,{\sc i} $2^3P_1 \to 2^3S_1$) provides an additional recombination
pathway at $z \gtrsim 1700$ and is included with its own Sobolev factor
and continuum opacity correction.

The matter temperature $T_m$ evolves via Compton heating/cooling against
the radiation field and adiabatic expansion:
\begin{equation}
\label{eq:Tmat}
\frac{dT_m}{dz} = \frac{8\sigma_T\,a_r\,T_r^4}{3\,H\,m_e\,c}
\frac{x_e}{1 + x_e + f_{\rm He}}(T_m - T_r) + \frac{2T_m}{1+z}\,,
\end{equation}
where $a_r = 4\sigma_{\rm SB}/c$ is the radiation constant.  At early
times the Compton timescale is much shorter than the Hubble time and
$T_m \approx T_r$; after decoupling, the matter cools adiabatically as
$T_m \propto (1+z)^2$~\cite{Scott:2009sz}.

\subsection{Reionisation}
\label{sec:reionisation}

At low redshift ($z \lesssim 10$), ultraviolet radiation from the first
stars and galaxies reionises the intergalactic medium.  We model this
using the CAMB parameterisation\footnote{\url{https://cosmologist.info/notes/CAMB.pdf}}: a $\tanh$ transition in $(1+z)^{3/2}$
centred on a reionisation redshift $z_{\rm re}$ with width $\delta z = 0.5$,
supplemented by a second helium reionisation at $z \approx 3.5$.

The reionisation redshift $z_{\rm re}$ is not specified directly; instead,
it is determined by requiring that the total reionisation optical depth
matches the input parameter $\tau_{\rm reion}$.  We solve this implicit
equation using Brent's method (\texttt{scipy.optimize.\allowbreak brentq}).

The final ionisation fraction is
$x_e(z) = \max[x_e^{\rm recomb}(z),\allowbreak x_e^{\rm reion}(z)]$,
ensuring a smooth transition from the recombination freeze-out to the
fully reionised state.

\subsection{Visibility function and derived quantities}
\label{sec:visibility}

The Thomson scattering opacity is
\begin{equation}
\label{eq:opacity}
\dot{\kappa}(\tau) = x_e\,n_H\,\sigma_T / a^2\,,
\end{equation}
where $n_H$ is the present-day hydrogen number density and $\sigma_T$ the
Thomson cross section.  The optical depth from conformal time $\tau$ to today is
\begin{equation}
\label{eq:optical_depth}
\tau_{\rm opt}(\tau) = \int_\tau^{\tau_0} \dot{\kappa}\,d\tau'\,,
\end{equation}
and the visibility function
\begin{equation}
\label{eq:visibility}
g(\tau) = \dot{\kappa}\,e^{-\tau_{\rm opt}}
\end{equation}
gives the probability that a CMB photon last scattered at conformal time $\tau$.
Its peak defines the surface of last scattering ($z_* \approx 1090$), and its
width $\Delta\tau_{\rm rec}$ determines the thickness of the last scattering surface.

The Silk damping scale, which sets the exponential suppression of small-scale
anisotropies due to photon diffusion, is
\begin{equation}
\label{eq:kD}
k_D^{-2} = \int_0^{a_*} \frac{R^2 + \frac{16}{15}(1+R)}{6(1+R)^2\,\dot{\kappa}}
\frac{da}{a^2 H}\,,
\end{equation}
where $R = 3\rho_b/(4\rho_\gamma)$ as before.

\begin{figure*}[!t]
\centering
\includegraphics[width=\textwidth]{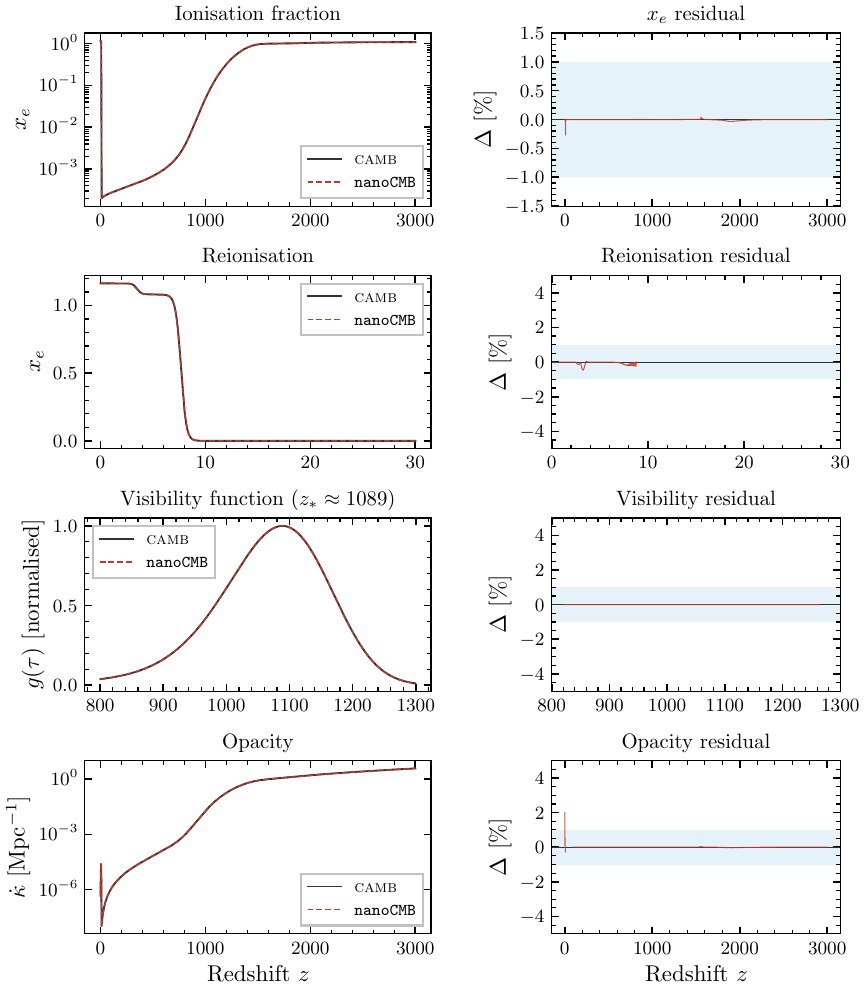}
\caption{\label{fig:thermo}
Thermodynamic quantities from \texttt{nanoCMB} (red dashed) compared to
CAMB (black solid).  {\em Top row:} ionisation fraction $x_e(z)$.
{\em Second row:} reionisation window at $z < 30$.
{\em Third row:} visibility function near recombination.
{\em Bottom row:} Thomson opacity $\dot{\kappa}$.
Right panels show ratios, demonstrating sub-percent agreement.}
\end{figure*}

Figure~\ref{fig:thermo} shows the ionisation history, visibility function,
and opacity from \texttt{nanoCMB} compared to CAMB.

\section{Perturbation theory}
\label{sec:perturbations}

\subsection{Synchronous gauge}
\label{sec:synchronous}

We work in the synchronous gauge with the CDM rest-frame convention, matching
CAMB~\cite{Lewis:1999bs} and following the notation of Ma \&
Bertschinger~\cite{Ma:1995ey}.  The perturbed metric is
\begin{equation}
ds^2 = a^2(\tau)\left[-d\tau^2 + (\delta_{ij} + h_{ij})\,dx^i\,dx^j\right]\,,
\end{equation}
where $h_{ij}$ decomposes into scalar modes characterised by $h$ and $\eta$
in Fourier space.  The key metric variable evolved in the code is
$k\eta$, the product of the wavenumber and the synchronous-gauge metric
perturbation $\eta$.

The Einstein constraint equations determine the gravitational potentials
from the matter perturbations:
\begin{align}
\label{eq:einstein_z}
\dot{h}/2 &= -k\,z\,,\quad z = \frac{\delta\rho/(2k) + k\eta}{\dot{a}/a}\,,\\
\label{eq:einstein_sigma}
\sigma &= z + \frac{3}{2}\frac{\delta q}{k^2}\,,
\end{align}
where $\delta\rho$ and $\delta q$ are the total density and momentum
perturbations, and $\sigma = (\dot{h} + 6\dot{\eta})/(2k)$ is the
shear.

\subsection{Boltzmann hierarchy}
\label{sec:hierarchy}

The distribution function for each species is expanded in Legendre multipoles.
We describe the evolution equations for each species in turn.

\subsubsection{Photons}
\label{sec:photons}

The photon density perturbation $\delta_\gamma = \Theta_0$ and momentum
perturbation $q_\gamma = (4/3)\Theta_1$ evolve as
\begin{align}
\dot{\delta}_\gamma &= -k\left(\tfrac{4}{3}\,z + q_\gamma\right)\,,\\
\dot{q}_\gamma &= \tfrac{k}{3}\left(\delta_\gamma - 2\,\pi_\gamma\right)
+ \dot{\kappa}(\tfrac{4}{3}\,v_b - q_\gamma)\,,
\end{align}
where $\pi_\gamma = \Theta_2$ is the photon anisotropic stress and the
Thomson scattering term in $\dot{q}_\gamma$ couples the photon and baryon
momenta.  The higher multipoles ($\ell \ge 2$) obey
\begin{multline}
\label{eq:photon_hierarchy}
\dot{\Theta}_\ell = \frac{k}{2\ell+1}\left[\ell\,\Theta_{\ell-1}
- (\ell+1)\,\Theta_{\ell+1}\right]
+ \delta_{\ell 2}\,\tfrac{8}{15}\,k\,\sigma \\
- \dot{\kappa}\left[\Theta_\ell - \delta_{\ell 0}\,\Theta_0
- \delta_{\ell 1}\,v_b - \delta_{\ell 2}\,\Pi\right]\,,
\end{multline}
where $\Pi = \Theta_2/10 + 3E_2/5$ is the polarisation source.

The photon polarisation hierarchy ($\ell \ge 2$) is
\begin{multline}
\label{eq:pol_hierarchy}
\dot{E}_\ell = \frac{k}{2\ell+1}\left[\ell\,E_{\ell-1}
- \frac{(\ell+3)(\ell-1)}{\ell+1}\,E_{\ell+1}\right] \\
- \dot{\kappa}\left[E_\ell - \delta_{\ell 2}\,\Pi\right]\,.
\end{multline}
Thomson scattering couples the temperature quadrupole $\Theta_2$ and
the polarisation quadrupole $E_2$ through $\Pi$, generating E-mode
polarisation from the photon anisotropy at the last scattering surface.

\subsubsection{Massless neutrinos}
\label{sec:neutrinos}

Massless neutrinos free-stream after decoupling at $T \sim 1\,$MeV and
obey an unscattered Boltzmann hierarchy.  The monopole and dipole are
\begin{align}
\dot{\delta}_\nu &= -k\left(\tfrac{4}{3}\,z + q_\nu\right)\,,\\
\dot{q}_\nu &= \tfrac{k}{3}\left(\delta_\nu - 2\,\pi_\nu\right)\,,
\end{align}
where $q_\nu = (4/3)\mathcal{N}_1$ is the neutrino momentum perturbation
and $\pi_\nu = \mathcal{N}_2$ the anisotropic stress.  The higher
multipoles ($\ell \ge 2$) obey
\begin{equation}
\label{eq:neutrino_hierarchy}
\dot{\mathcal{N}}_\ell = \frac{k}{2\ell+1}\left[\ell\,\mathcal{N}_{\ell-1}
- (\ell+1)\,\mathcal{N}_{\ell+1}\right]
+ \delta_{\ell 2}\,\tfrac{8}{15}\,k\,\sigma\,.
\end{equation}
The $\ell = 2$ source term from the metric shear $\sigma$ seeds the
neutrino anisotropic stress, which in turn affects the gravitational
potentials through the Einstein equations.  Unlike photons, there is no
scattering term, so all multipoles are populated by free streaming.

\subsubsection{Cold dark matter}
\label{sec:cdm}

In the synchronous gauge with the CDM at rest, the CDM equations are
simply
\begin{equation}
\dot{\delta}_c = -k\,z\,,
\end{equation}
and the CDM velocity is zero by gauge choice.  CDM is pressureless and
collisionless, so it has no higher multipoles.

\subsubsection{Baryons}
\label{sec:baryons}

The baryon density and velocity evolve as
\begin{align}
\dot{\delta}_b &= -k(z + v_b)\,,\\
\dot{v}_b &= -\frac{\dot{a}}{a}\,v_b + k\,c_{s,b}^2\,\delta_b
- \frac{\dot{\kappa}\,\bar{\rho}_\gamma}{a\,\bar{\rho}_b}\left(\tfrac{4}{3}\,v_b - q_\gamma\right)\,,
\end{align}
where the last term in the velocity equation is the Thomson drag force
coupling baryons to photons.  The baryon sound speed is
\begin{equation}
\label{eq:cs2b}
c_{s,b}^2 = \frac{k_B\,T_m}{\mu\,m_H}\left(1 - \frac{1}{3}\frac{d\ln T_m}{d\ln a}\right)\,,
\end{equation}
with $\mu^{-1} = 1 - \tfrac{3}{4}Y_{\rm He} + (1 - Y_{\rm He})\,x_e$ the inverse
mean molecular weight in units of $m_H$.  At early times, Compton scattering
keeps $T_m \approx T_r \propto 1/a$ so that $c_{s,b}^2 = (4/3)\,k_B T_m / (\mu\,m_H)$;
after decoupling, $T_m \propto a^{-2}$ (adiabatic cooling) and the factor approaches
$5/3$.  This thermal pressure term is a small correction but is included for
consistency with CAMB.

The state vector is laid out as a flat array for the SciPy ODE solver:

\begin{lstlisting}[style=nanoCMB, caption={State vector index layout.}]
LMAXG = 15        # photon temperature: Theta_0 ... Theta_LMAXG
LMAXPOL = 15      # photon polarisation: E_2 ... E_LMAXPOL
LMAXNR = 15       # massless neutrinos: N_0 ... N_LMAXNR

IX_ETAK = 0                            # k*eta (metric)
IX_CLXC = 1                            # delta_c (CDM density)
IX_CLXB = 2                            # delta_b (baryon density)
IX_VB   = 3                            # v_b (baryon velocity)
IX_G    = 4                            # Theta_0, Theta_1, ...
IX_POL  = IX_G + LMAXG + 1            # E_2, E_3, ...
IX_R    = IX_POL + LMAXPOL - 1         # N_0, N_1, ...
NVAR    = IX_R + LMAXNR + 1            # total: 50 variables
\end{lstlisting}

\subsection{Tight-coupling approximation}
\label{sec:tightcoupling}

At early times when the photon mean free path is much smaller than the
wavelength of the perturbation ($k/\dot{\kappa} \ll 1$) and the Hubble time
($1/(\dot{\kappa}\tau) \ll 1$), the photon--baryon system is tightly coupled
and the higher photon multipoles are exponentially suppressed.  In this regime,
evolving the full hierarchy is both unnecessary and numerically stiff.

In tight coupling, the photon--baryon system is treated as a single fluid,
with the photon quadrupole determined algebraically from the shear:
\begin{equation}
\label{eq:pig_tc}
\Theta_2^{\rm TC} = \frac{32}{45}\frac{k}{\dot{\kappa}}\left(\sigma + v_b\right)\,,
\end{equation}
and the baryon velocity equation becomes
\begin{equation}
\label{eq:vb_tc}
\dot{v}_b = \frac{-\frac{\dot{a}}{a}\,v_b + k\,c_{s,b}^2\,\delta_b + \frac{k}{4}\,\frac{4\rho_\gamma}{3\rho_b}\left(\Theta_0 - 2\Theta_2^{\rm TC}\right)}{1 + \frac{4\rho_\gamma}{3\rho_b}}\,.
\end{equation}

This approximation reduces the stiffness of the system and allows the ODE
solver to take much larger time steps during the tightly-coupled era.  The
transition criterion is $\max(k/\dot{\kappa},\, 1/(\dot{\kappa}\tau)) < 0.01$.

\subsection{Adiabatic initial conditions}
\label{sec:ics}

The perturbations are initialised deep in the radiation era when $k\tau \ll 1$,
using the standard adiabatic growing mode~\cite{Ma:1995ey}.  The
leading-order terms are:
\begin{align}
\label{eq:ic_etak}
k\eta &= -k\left(1 - \frac{(k\tau)^2}{12}\left(1 - \frac{10}{4R_\nu+15}\right)\right)\,,\\
\delta_\gamma = \delta_\nu &= \frac{(k\tau)^2}{3}\,,\qquad
\delta_c = \delta_b = \frac{3}{4}\,\delta_\gamma\,,
\end{align}
where $R_\nu = \rho_\nu/(\rho_\nu + \rho_\gamma)$ is the neutrino fraction
of the radiation density.  The photon and baryon velocities start at
$(3/4)q_\gamma$, and the neutrino anisotropic stress is seeded at second
order in $k\tau$:
\begin{equation}
\pi_\nu = -\frac{4}{3}\frac{(k\tau)^2}{4R_\nu + 15}\,.
\end{equation}

\begin{lstlisting}[style=nanoCMB, caption={Adiabatic initial conditions (excerpt).}]
x = k * tau;  x2 = x * x
Rv = bg["grhornomass"] / grho_rad     # neutrino fraction
Rp15 = 4 * Rv + 15

y0[IX_ETAK] = -k * (1.0 - x2/12.0 * (-10.0/Rp15 + 1.0))
clxg_init = x2 / 3.0
y0[IX_G]     = clxg_init              # delta_gamma
y0[IX_CLXC]  = 0.75 * clxg_init      # delta_c = (3/4)*delta_gamma
y0[IX_CLXB]  = 0.75 * clxg_init      # delta_b
y0[IX_R]     = clxg_init              # delta_nu = delta_gamma
y0[IX_R + 2] = -4.0/3.0 * x2 / Rp15  # pi_nu
\end{lstlisting}

\subsection{Hierarchy truncation}
\label{sec:truncation}

The Boltzmann hierarchy is formally infinite.  We truncate at
$\ell_{\max} = 15$ for all species, using the free-streaming closure
relation for the highest multipole:
\begin{equation}
\label{eq:truncation}
\dot{\Theta}_{\ell_{\max}} = k\,\Theta_{\ell_{\max}-1}
- \frac{\ell_{\max}+1}{\tau}\,\Theta_{\ell_{\max}}
- \dot{\kappa}\,\Theta_{\ell_{\max}}\,,
\end{equation}
where the $(\ell_{\max}+1)/\tau$ term approximates the coupling to the
unresolved $\ell_{\max}+1$ mode using the free-streaming relation for a
plane wave.  The same closure is used for neutrinos (without the scattering
term).

The perturbation evolution is carried out using SciPy's LSODA integrator,
which automatically switches between the non-stiff Adams method and the
stiff BDF method as needed.  This is well-suited to the transition from
tight coupling to free streaming.

\begin{figure*}[!t]
\centering
\includegraphics[width=\textwidth]{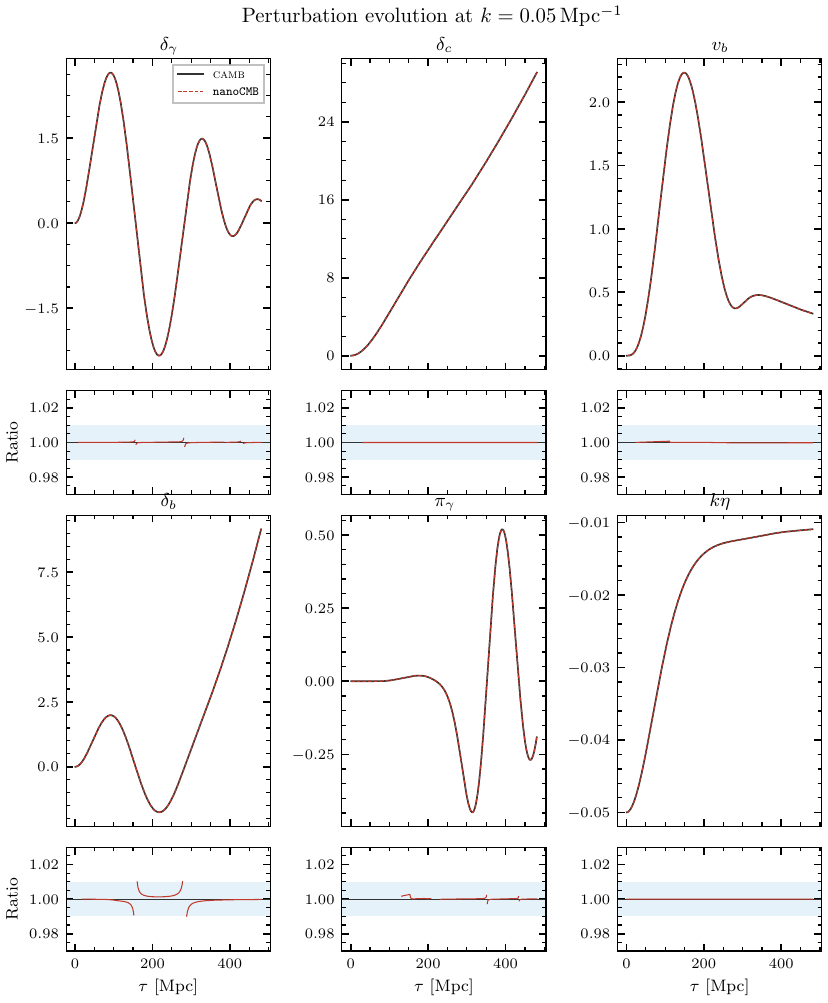}
\caption{\label{fig:perturbations}
Perturbation evolution at $k = 0.05\,$Mpc$^{-1}$.  Top panels show
$\delta_\gamma$, $\delta_c$, $v_b$, $\delta_b$, $\pi_\gamma$, and
$k\eta$ from \texttt{nanoCMB} (red dashed) and CAMB (black solid).
Bottom panels show ratios, with the $\pm$1\% band shaded.}
\end{figure*}

Figure~\ref{fig:perturbations} shows the evolution of all perturbation
variables at $k = 0.05\,$Mpc$^{-1}$, compared to CAMB.

\section{Source functions and line-of-sight integration}
\label{sec:los}

\subsection{The formal line-of-sight integral}
\label{sec:formal_los}

Rather than tracking the full photon distribution to each multipole $\ell$
up to $\ell_{\max} \sim 2500$, we use the line-of-sight (LOS) integration
method of Seljak \& Zaldarriaga~\cite{Seljak:1996is}.  The transfer
function at multipole $\ell$ for wavenumber $k$ is
\begin{equation}
\label{eq:los}
\Delta_\ell^T(k) = \int_0^{\tau_0} S_T(k, \tau)\,j_\ell(k\chi)\,d\tau\,,
\end{equation}
where $\chi = \tau_0 - \tau$ is the comoving distance and $j_\ell$ is the
spherical Bessel function.  The source function $S_T$ encodes the photon
monopole, dipole, and quadrupole contributions, weighted by the visibility
function and its derivatives.

\subsection{Integration by parts decomposition}
\label{sec:ibp}

The raw source function involves derivatives of the visibility function
$g(\tau)$, which are numerically noisy.  CAMB handles this by computing
the visibility derivatives $\dot{g}$ and $\ddot{g}$ directly and folding
them into a single composite source multiplied by $j_\ell$.  We take the
alternative approach of integrating by parts to transfer these derivatives
onto the smooth Bessel functions, decomposing the temperature transfer into
three channels:
\begin{equation}
\label{eq:ibp}
\Delta_\ell^T(k) = \int d\tau\left[
S_0\,j_\ell(k\chi) + S_1\,j_\ell'(k\chi) + S_2\,j_\ell''(k\chi)
\right]\,,
\end{equation}
where the three source terms are:
\begin{align}
\label{eq:src_j0}
S_0 &= 2\dot{\Phi}\,e^{-\tau_{\rm opt}} \notag\\
    &\quad + g\!\left(-\eta + 2\Phi + \tfrac{\Theta_0}{4} + \tfrac{5}{8}\Pi\right)\,,\\
\label{eq:src_j1}
S_1 &= g\left(\sigma + v_b\right)\,,\\
\label{eq:src_j2}
S_2 &= \tfrac{15}{8}\,g\,\Pi\,.
\end{align}

The gravitational potential $\Phi$ is determined by the Einstein constraint
\begin{equation}
\label{eq:phi}
\Phi = -\frac{\delta\rho + 3(\dot{a}/a)\,\delta q/k + \delta\pi}{2k^2}\,,
\end{equation}
and its time derivative $\dot{\Phi}$ (the ISW source) is computed from the
perturbed Einstein equations.

The E-mode polarisation transfer function uses a simpler form:
\begin{equation}
\label{eq:los_E}
\Delta_\ell^E(k) = \int d\tau\;\frac{15}{8}\frac{g\,\Pi}{\chi^2 k^2}\;j_\ell(k\chi)\,.
\end{equation}

In the code, the three temperature source channels are assembled as:

\begin{lstlisting}[style=nanoCMB, caption={Source function assembly (IBP decomposition).}]
src_j0 = ISW_arr + vis_arr * (monopole_arr + 0.625 * polter_arr)
src_j1 = vis_arr * sigma_plus_vb_arr
src_j2 = 1.875 * vis_arr * polter_arr
\end{lstlisting}

\subsection{Gravitational potentials and the ISW effect}
\label{sec:isw}

The integrated Sachs--Wolfe (ISW) effect arises from the time derivative of
the gravitational potential $\dot{\Phi}$.  This is non-zero during the
radiation-to-matter transition (early ISW, which boosts the first acoustic
peak) and during dark energy domination at late times (late ISW, which
affects low $\ell$).

Computing $\dot{\Phi}$ requires the time derivatives of the photon and
neutrino anisotropic stresses $\dot{\pi}_\gamma$ and $\dot{\pi}_\nu$,
which we obtain directly from the Boltzmann hierarchy equations rather
than by numerical differentiation:
\begin{multline}
\dot{\Phi} = \frac{1}{2k^2}\bigg[\frac{\dot{a}}{a}\!\left(-\delta\pi - 2k^2\Phi\right)
+ k\,\delta q \\
- \dot{\delta\pi}
+ k\sigma(\bar{\rho}_{\rm tot} + \bar{P}_{\rm tot})\bigg]\,,
\end{multline}
where $\dot{\delta\pi} = \bar{\rho}_\gamma\,\dot{\pi}_\gamma + \bar{\rho}_\nu\,\dot{\pi}_\nu
- 4(\dot{a}/a)\,\delta\pi$.

\subsection{Precomputed Bessel function tables}
\label{sec:bessel}

The spherical Bessel functions $j_\ell(x)$ are needed at millions of points
during the LOS integration.  We precompute them on a uniform $x$-grid with
spacing $\Delta x = 0.03$, then interpolate linearly during integration.
Several optimisations are employed:
\begin{itemize}
\item {\em Dead-zone skipping:} For large $\ell$, $j_\ell(x)$ is
exponentially small for $x < \ell - O(\ell^{1/3})$.  We skip this region
during both evaluation and integration.
\item {\em Bessel derivatives by recurrence:} Rather than computing
$j_\ell'$ and $j_\ell''$ independently, we use the standard recurrence
relations:
\begin{align}
\label{eq:bessel_deriv}
j_\ell'(x) &= \frac{\ell}{x}\,j_\ell(x) - j_{\ell+1}(x)\,,\\
j_\ell''(x) &= -\frac{2}{x}\,j_\ell'(x) + \left[\frac{\ell(\ell+1)}{x^2} - 1\right]j_\ell(x)\,,
\end{align}
which require only $j_\ell$ and $j_{\ell+1}$ (both already tabulated).
\item {\em Disk caching:} Bessel tables are cached to disk (keyed by a hash
of the $\ell$-values and grid parameters), so subsequent runs skip the
expensive evaluation entirely.
\end{itemize}

\subsection{Practical LOS integration}
\label{sec:practical_los}

For each multipole $\ell$, we restrict the $k$-integration range to where
$j_\ell(k\chi)$ is non-negligible.  Specifically, we integrate from
$k_{\rm lo} = (\ell - 4\ell^{1/3})/\chi_{\max}$ to
$k_{\rm hi} = (\ell + 2500)/\chi_*$, where $\chi_* = \tau_0 - \tau_*$.
Outside this range, the Bessel function is either in the dead zone or has
oscillated to negligible amplitude after averaging.

The LOS integrals for different $\ell$ values are independent and are
computed in parallel using Python's \texttt{Thread\-Pool\-Executor}.  This
provides near-linear speedup on multi-core machines since the computation
is dominated by NumPy array operations that release the GIL.

\begin{figure*}[!t]
\centering
\includegraphics[width=\textwidth]{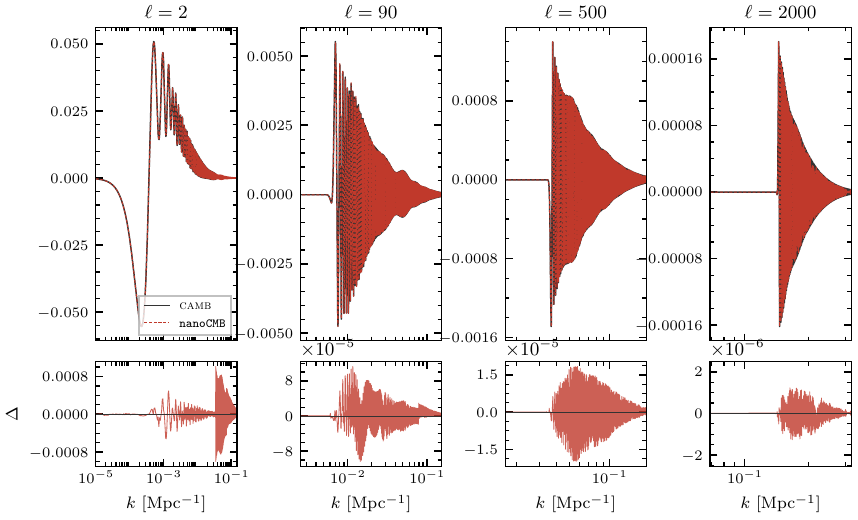}
\caption{\label{fig:transfer}
Temperature transfer functions $\Delta_\ell^T(k)$ at four representative
multipoles.  Top panels show \texttt{nanoCMB} (red dashed) and CAMB
(black solid); bottom panels show residuals.}
\end{figure*}

Figure~\ref{fig:transfer} compares the resulting temperature transfer functions
$\Delta_\ell^T(k)$ at four representative multipoles spanning
$\ell = 2$ to $\ell = 2000$.  The oscillatory structure from the Bessel
functions is well reproduced, and residuals are at the sub-percent level
across the range of $k$ that contributes significantly to $C_\ell$.

\section{Power spectrum assembly}
\label{sec:cls}

\subsection{From transfer functions to angular power spectra}
\label{sec:cl_integral}

The angular power spectrum is
\begin{equation}
\label{eq:cl}
C_\ell^{XY} = 4\pi \int d(\ln k)\;\mathcal{P}(k)\;\Delta_\ell^X(k)\,\Delta_\ell^Y(k)\,,
\end{equation}
where $\mathcal{P}(k) = A_s\,(k/k_0)^{n_s-1}$ is the dimensionless
primordial power spectrum and $X, Y \in \{T, E\}$.

The conventional plotting quantity is $\mathcal{D}_\ell = \ell(\ell+1)C_\ell/(2\pi)$
in units of $\mu$K$^2$.  For E-mode polarisation, an additional normalisation
factor of $[\ell(\ell-1)(\ell+1)(\ell+2)]^{1/2}$ per transfer function
accounts for the spin-2 nature of the polarisation field.  For the cross-spectrum
$C_\ell^{TE}$, this factor enters as the square root.

In code, the assembly is:

\begin{lstlisting}[style=nanoCMB, caption={Power spectrum assembly.}]
Pk = A_s * (k_fine / k_pivot)**(n_s - 1.0)

Cl_TT, Cl_EE, Cl_TE = [np.trapezoid(Pk * d, lnk_fine, axis=1)
    for d in (Delta_T**2, Delta_E**2, Delta_T * Delta_E)]

norm = 4.0 * np.pi * ells * (ells + 1) / (2.0 * np.pi)
ctnorm = (ells**2 - 1.0) * (ells + 2) * ells   # E-mode factor
Cl_TT *= norm;  Cl_EE *= norm * ctnorm
Cl_TE *= norm * np.sqrt(ctnorm)
\end{lstlisting}

\subsection{Two-grid strategy}
\label{sec:twogrid}

The source functions $S(k, \tau)$ vary smoothly with $k$ and can be
accurately computed on a relatively coarse ODE grid.
However, the transfer functions $\Delta_\ell(k)$ oscillate rapidly (due to
Bessel function ringing), and the $C_\ell$ integrand
$|\Delta_\ell(k)|^2$ requires much finer $k$-sampling.

Our strategy is:
\begin{enumerate}
\item Solve the ODE system on 200 non-uniform $k$-modes (section~\ref{sec:grids}).
\item Interpolate the source functions to a finer grid of 4000 $k$-modes using
Akima interpolation~\cite{Akima1970}, which provides smooth results without
the overshooting artefacts of cubic splines.
\item Perform the LOS integral and $C_\ell$ integration on the fine grid.
\end{enumerate}

This two-grid approach saves computation by a factor of $\sim$10 compared
to solving the ODE on the fine grid directly.

\subsection{Multipole sampling and interpolation}
\label{sec:ell_sampling}

Rather than computing $C_\ell$ at every integer $\ell$ from 2 to 2500, we
evaluate at a sparse set of $\sim$100 multipoles: every other $\ell$ for
$\ell < 40$, every 5th for $40 \le \ell < 200$, and every 50th for
$\ell \ge 200$.  The full spectrum is then obtained by cubic spline
interpolation in $\ell$.  This is accurate because $C_\ell$ varies
smoothly with $\ell$ (the rapid oscillations are in $k$, not $\ell$).

\section{Optimal non-uniform grids}
\label{sec:grids}
\subsection{Error equidistribution principle}
\label{sec:equidistribution}

The accuracy of the trapezoidal rule applied to $\int f(x)\,dx$ on a non-uniform grid
$\{x_0, x_1, \ldots, x_N\}$ is controlled by the local error
$\epsilon_i \sim h_i^3\,|f''(x_i)|$, where $h_i = x_{i+1} - x_i$.  For a
fixed total number of points $N$, the global error is minimised when the
local errors are equidistributed across all intervals.  This gives the
optimal node density
\begin{equation}
\label{eq:equidist}
\frac{dN}{dx} \propto |f''(x)|^{1/3}\,.
\end{equation}

In practice, the integrand is not known a priori --- its computation is
the goal of the grid construction --- so we cannot evaluate $|f''|$
directly.  The equidistribution principle nevertheless motivates the
general strategy: construct a weight function $w(x)$ that is large where
the integrand is expected to have the most structure (large amplitude and
rapid oscillation), then place nodes according to the cumulative
distribution of $w^{1/3}$.  

\subsection{Wavenumber grid}
\label{sec:kgrid}

\begin{figure*}[!t]
\centering
\includegraphics[width=\textwidth]{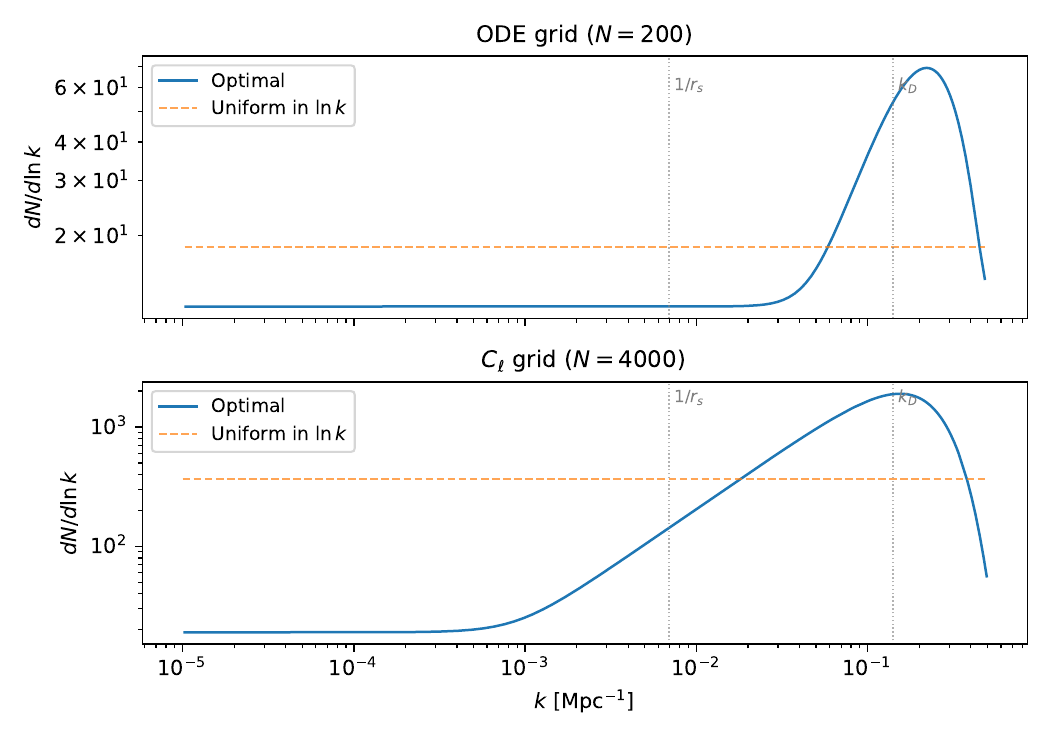}\\[0.5em]
\includegraphics[width=0.6\textwidth]{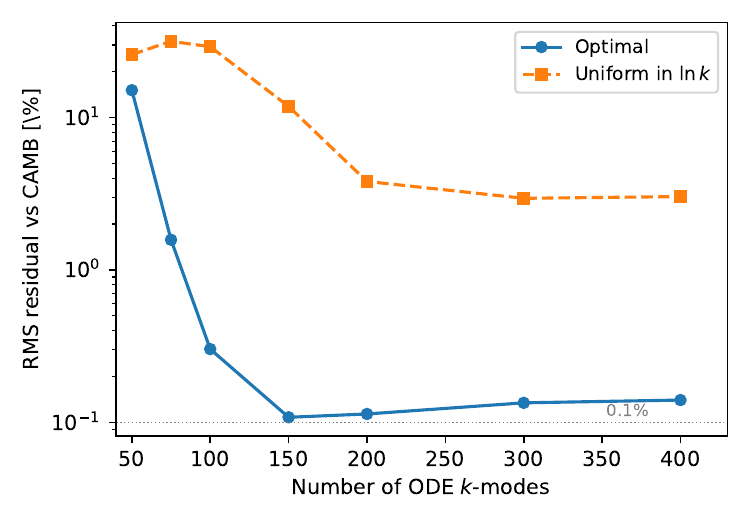}
\caption{\label{fig:kgrid}
{\em Top panels:} Node density $dN/d\ln k$ for the optimal (blue solid)
and uniform-in-$\ln k$ (orange dashed) grids, for the ODE grid
($N = 200$, top) and $C_\ell$ grid ($N = 4000$, bottom).  Vertical
dotted lines mark the acoustic scale $1/r_s$ and the Silk damping
scale $k_D$.
{\em Bottom panel:} $TT$ RMS residual versus CAMB as a function of
the number of ODE $k$-modes, showing the convergence advantage of
the optimal grid.}
\end{figure*}

A na\"ive choice is to distribute $k$-modes uniformly in $\ln k$, but this
wastes points in regions where the integrand is smooth (low $k$) while
under-resolving the acoustic oscillations near the damping scale.  A physically motivated
weight function concentrates points where they matter most, achieving the
same accuracy with significantly fewer modes (figure~\ref{fig:kgrid}).

We construct two $k$-grids using the density~\eqref{eq:equidist}:

\paragraph{ODE grid ($N = 200$):} The weight function captures the acoustic
oscillation structure and damping of the source functions:
\begin{equation}
\label{eq:wode}
w_{\rm ode}(k) = \underbrace{\max\left(\frac{1}{r_s^2},\, \frac{(k\,r_s)^2}{\tau_{\rm eq}^2}\right)}_{\text{curvature}}
\times \underbrace{k^{n_s+2}}_{\text{primordial}} \times\; \underbrace{e^{-(k/k_D)^2}}_{\text{Silk damping}}\,.
\end{equation}

Each factor in this weight function has a direct physical motivation:

\begin{itemize}

\item {\em Silk damping:} The exponential $e^{-(k/k_D)^2}$ reflects the
suppression of the source functions by photon diffusion during recombination.
Perturbation modes with $k \gg k_D$ are exponentially damped, so there is
no benefit in placing ODE grid points in this region.  

\item {\em Primordial weighting:} The factor $k^{n_s+2}$ combines three
contributions: the primordial power spectrum $\mathcal{P}(k) \propto k^{n_s-1}$,
the logarithmic integration measure $d\ln k$ (contributing one power of $k$),
and an additional factor of $k^2$ reflecting the phase-space weighting of
higher-$k$ modes in the $C_\ell$ integral.  Together, these determine the
{\em amplitude} of the integrand at each $k$ before accounting for oscillatory
structure.

\item {\em Curvature estimate:} The first factor determines where
interpolation errors are largest by estimating how rapidly the source
functions oscillate.  Acoustic waves have period $\sim\!\pi/r_s$ in
$k$-space, so the second derivative of the oscillatory structure scales
as $(k\,r_s)^2$ --- this is the one factor with a direct connection to
$|f''|$.  Dividing by $\tau_{\rm eq}^2$ normalises this against the
timescale over which the transfer function envelope varies, giving a
dimensionless measure of the curvature in the acoustic regime.  The
$\max$ with $1/r_s^2$ provides a floor for super-horizon modes
($k\,r_s \ll 1$), where there are no acoustic oscillations and the source
functions are smooth, but some grid points are still needed to resolve the
Sachs--Wolfe plateau.  The transition between the two branches occurs near
$k \sim 1/\tau_{\rm eq}$, which corresponds approximately to the scale
entering the horizon at matter--radiation equality --- the boundary between
modes that experienced acoustic oscillations and those that did not.

\end{itemize}

\paragraph{$C_\ell$ grid ($N = 4000$):} The $C_\ell$ integrand
$\mathcal{P}(k)\,|\Delta_\ell(k)|^2$ oscillates rapidly in $k$ due to
Bessel function ringing, requiring much finer sampling than the ODE grid.
The weight function accounts for the Bessel function windowing that selects
$k \approx \ell/\chi_*$ for each multipole:
\begin{multline}
w_{\rm cl}(k) = \sum_\ell \exp\!\left[-\frac{(k - \ell/\chi_*)^2}{2(3\sigma_k)^2}\right] \\
\times \max\!\left(\frac{1}{r_s^2},\, \sigma_k^2\right)
k^{n_s+2}\,e^{-(k/k_D)^2}\,,
\end{multline}
where $\sigma_k = 1/\Delta\tau_{\rm rec}$ is the width of the visibility
function in $k$-space.  The Gaussian envelope centred on $k = \ell/\chi_*$
approximates the Bessel window through which each multipole samples the
source function; the sum over $\sim$30 geometrically spaced $\ell$ values
ensures that the grid resolves the integrand across the full multipole range.
The curvature factor now uses $\max(1/r_s^2,\, \sigma_k^2)$: in the acoustic
regime, the Bessel oscillation period $\sim\!1/\Delta\tau_{\rm rec}$ sets the
dominant curvature scale rather than the acoustic scale $1/r_s$, since the
$C_\ell$ integrand oscillates at the Bessel frequency, not the acoustic
frequency.

\begin{lstlisting}[style=nanoCMB, caption={$k$-grid density construction (ODE mode).}]
# Source-level damping: exp(-(k/k_D)^2)
primordial = k ** (params["n_s"] + 2)
damped = primordial * np.exp(-1.0 * (k / thermo["k_D"]) ** 2)

# Acoustic curvature
smooth_curv = (k * thermo["r_s"]) ** 2 / bg["tau_eq"] ** 2
raw_weight = np.maximum(1/thermo["r_s"]**2, smooth_curv) * damped

# Node density ~ weight^(1/3), with 0.5% floor
density = (raw_weight + 0.005 * np.max(raw_weight)) ** (1.0/3.0)
\end{lstlisting}

Figure~\ref{fig:kgrid} (top panels) shows the resulting node density
$dN/d\ln k$ for both grids compared to uniform $\ln k$ spacing.  The optimal
grids concentrate points in the acoustic regime ($k \sim 1/r_s$ to
$k \sim k_D$) while the uniform grid spreads them evenly.
Figure~\ref{fig:kgrid} (bottom panel) demonstrates the payoff: the $TT$
RMS residual versus CAMB as a function of the number of ODE $k$-modes,
comparing the optimal and uniform grids.  The optimal grid reaches
$\sim$0.1\% accuracy with $\sim$150 modes, whereas the uniform grid
saturates at $\sim$3\% even at $N = 400$ because the evenly spread nodes
cannot resolve the acoustic oscillation structure near the damping scale.

\subsection{Conformal time grid}
\label{sec:taugrid}

The $\tau$-grid ($N = 1000$) must resolve three features: the narrow
visibility peak at $\tau_*$, the broader acoustic source structure (width
$\sim r_s$), and the reionisation bump at $\tau_{\rm reion}$:
\begin{multline}
w_\tau(\tau) = \frac{1}{\Delta\tau_{\rm rec}^2}\,e^{-(\tau-\tau_*)^2/(2\Delta\tau_{\rm rec}^2)}
+ \frac{k_{\max}^2}{3}\,e^{-(\tau-\tau_*)^2/(2r_s^2)} \\
+ \frac{0.3}{\Delta\tau_{\rm reion}^2}\,e^{-(\tau-\tau_{\rm reion})^2/(2\Delta\tau_{\rm reion}^2)}\,.
\end{multline}

The node density is again $\propto w^{1/3}$, with a 0.5\% floor.
This concentrates points where the integrand has the most structure while
maintaining adequate coverage elsewhere.

\section{Validation}
\label{sec:validation}

All comparisons in this section (and in the preceding figures) use
CAMB~\cite{Lewis:1999bs} run with \texttt{AccuracyBoost=3}, configured with lensing
disabled, no tensors, and massless neutrinos only, matching the
\texttt{nanoCMB} assumptions.  Figures use the Planck 2018 best-fit
parameters listed in table~\ref{tab:params}; the multi-cosmology
benchmark (section~\ref{sec:benchmark}) spans $\pm 3\sigma$ of the
Planck posterior.

As shown in the preceding sections, the intermediate quantities agree well
with CAMB: the Hubble parameter to $\sim 10^{-4}$ (figure~\ref{fig:background}),
the ionisation history and visibility function to $\lesssim 0.5\%$
(figure~\ref{fig:thermo}), the perturbation variables to $\sim 1\%$
at all $k$-modes tested (figure~\ref{fig:perturbations}), and the transfer
functions to sub-percent accuracy (figure~\ref{fig:transfer}).

\subsection{Power spectra}
\label{sec:val_cls}

\begin{figure*}[!t]
\centering
\includegraphics[width=\textwidth]{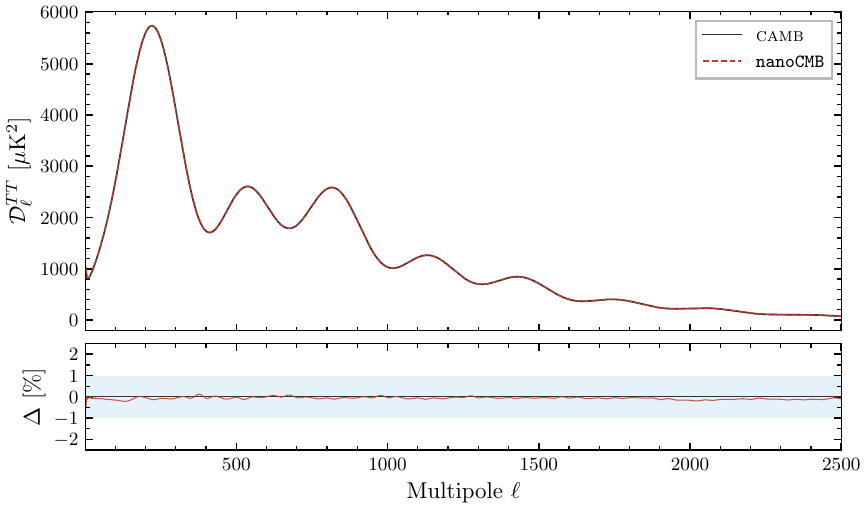}
\caption{\label{fig:spectra_tt}
Temperature power spectrum $\mathcal{D}_\ell^{TT}$ from \texttt{nanoCMB}
(red) compared to CAMB (black).  The lower panel shows residuals as
percentages, with the $\pm 1\%$ band shaded.}
\end{figure*}

\begin{figure*}[!t]
\centering
\includegraphics[width=\textwidth]{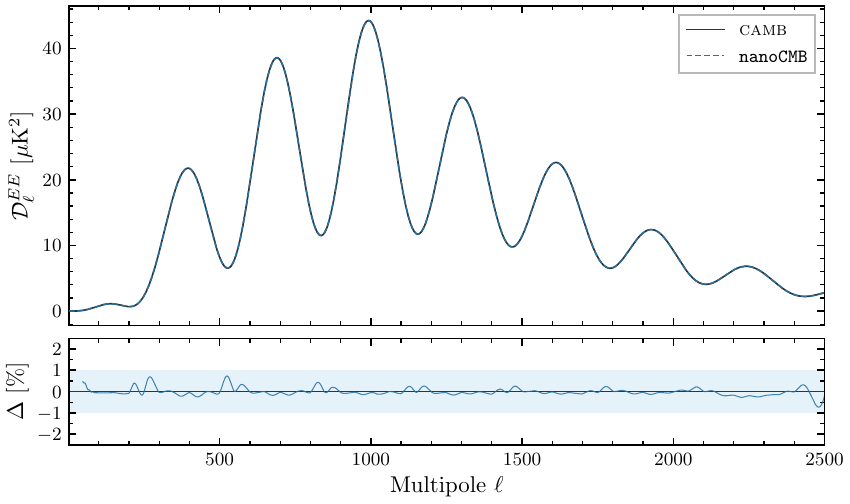}\\[0.5em]
\includegraphics[width=\textwidth]{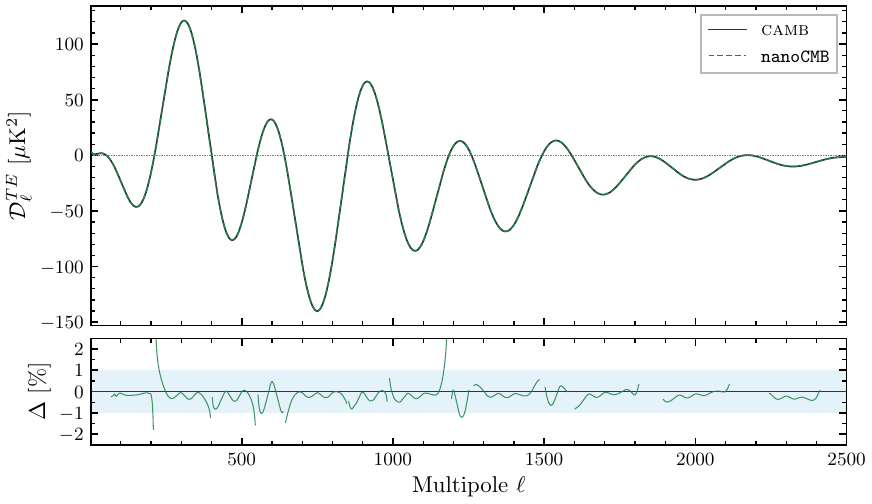}
\caption{\label{fig:spectra_ee_te}
E-mode polarisation ($\mathcal{D}_\ell^{EE}$, top) and temperature--polarisation
cross-spectrum ($\mathcal{D}_\ell^{TE}$, bottom) from \texttt{nanoCMB}
compared to CAMB.  Lower panels show residuals with the $\pm 1\%$ band shaded.}
\end{figure*}

Figures~\ref{fig:spectra_tt} and~\ref{fig:spectra_ee_te} show the
$TT$, $EE$, and $TE$ angular power spectra from \texttt{nanoCMB}
compared to CAMB for the fiducial parameters, with residual panels.
Table~\ref{tab:accuracy} summarises the accuracy for this fiducial
cosmology.

\begin{table}[tbp]
\centering
\begin{tabular}{lcccc}
\toprule
& \multicolumn{2}{c}{$TT$} & \multicolumn{2}{c}{$EE$} \\
\cmidrule(lr){2-3}\cmidrule(lr){4-5}
$\ell$ range & Mean ratio & Std & Mean ratio & Std \\
\midrule
$[2, 29]$     & 0.9993 & 0.0008 & 1.0013 & 0.0099 \\
$[30, 499]$   & 0.9996 & 0.0009 & 1.0005 & 0.0022 \\
$[500, 1999]$ & 0.9998 & 0.0008 & 1.0003 & 0.0013 \\
$[2000, 2500]$& 0.9982 & 0.0005 & 0.9986 & 0.0020 \\
\bottomrule
\end{tabular}
\caption{\label{tab:accuracy}
Accuracy summary for the fiducial cosmology: mean ratio and standard deviation of
the nanoCMB/CAMB power spectra by multipole range. Sub-percent accuracy is
achieved across the full range $2 \le \ell \le 2500$.}
\end{table}

\subsection{Multi-cosmology benchmark}
\label{sec:benchmark}

To verify that the accuracy is not tuned to a single parameter set, we
benchmark \texttt{nanoCMB} against CAMB across 50 flat $\Lambda$CDM
cosmologies drawn from a Latin hypercube spanning $\pm 3\sigma$ of the
Planck 2018 posterior~\cite{Planck:2018vyg}, varying $\omega_b$, $\omega_c$,
$h$, $n_s$, $\ln(10^{10}A_s)$, and $\tau_{\rm reion}$ simultaneously.
All 50 runs completed successfully.

Table~\ref{tab:benchmark} reports the RMS fractional residual
\[
\left[\left\langle\!\left(\frac{\mathcal{D}_\ell^{\rm nano}}{\mathcal{D}_\ell^{\rm CAMB}} - 1\right)^{\!2}
\right\rangle_{\!\ell}\right]^{1/2}
\]
within each multipole bin, summarised as the
median and 95th percentile across the 50 cosmologies.  For $TT$, the
RMS residual is $\lesssim$0.1\% at $\ell < 2000$ and $\sim$0.2\% in the
highest bin, with negligible variation across cosmologies.  For $EE$ at
$\ell \ge 30$, the RMS residual is 0.1--0.3\%, comparable to $TT$.

The larger RMS for $EE$ at $\ell < 30$ reflects the very small amplitude
of the reionisation bump ($\mathcal{D}_\ell^{EE} \lesssim 0.04\;\mu\text{K}^2$),
where even sub-$\mu$K$^2$ absolute differences produce sizeable
fractional residuals.  We do not report ratio-based statistics for
$C_\ell^{TE}$ because the cross-spectrum passes through zero at multiple
multipoles, making fractional residuals ill-defined; the $TE$ residuals
visible in figure~\ref{fig:spectra_ee_te} are representative.

\begin{table}[tbp]
\centering
\begin{tabular}{lcccc}
\toprule
& \multicolumn{2}{c}{$TT$} & \multicolumn{2}{c}{$EE$} \\
\cmidrule(lr){2-3}\cmidrule(lr){4-5}
$\ell$ range & Median & 95th \%ile & Median & 95th \%ile \\
\midrule
$[2, 29]$      & 0.10\% & 0.12\% & 1.0\%$^\dagger$ & 4.4\%$^\dagger$ \\
$[30, 499]$    & 0.09\% & 0.10\% & 0.22\% & 0.24\% \\
$[500, 1999]$  & 0.08\% & 0.08\% & 0.14\% & 0.15\% \\
$[2000, 2500]$ & 0.19\% & 0.21\% & 0.24\% & 0.29\% \\
\bottomrule
\end{tabular}
\caption{\label{tab:benchmark}
Multi-cosmology benchmark: median and 95th-percentile of the RMS
fractional residual in each multipole bin, across 50 cosmologies
sampled over $\pm 3\sigma$ of the Planck 2018 posterior.
$^\dagger$Elevated at low $\ell$ due to the small amplitude of the
$EE$ reionisation bump.}
\end{table}

To put these residuals in context, CAMB's internal convergence at default
accuracy is $\sim$0.05--0.1\% RMS when comparing \texttt{AccuracyBoost=1}
to \texttt{AccuracyBoost=4}, comparable to the \texttt{nanoCMB} residuals.
In other words, \texttt{nanoCMB} achieves accuracy comparable to CAMB at
its default settings.

\subsection{Timing}
\label{sec:timing}

On an Apple M-series laptop (12 cores), the full calculation takes
approximately 30\,s without Numba, dominated by the perturbation evolution
of 200 $k$-modes parallelised across cores.  With Numba JIT compilation
enabled (optional), the runtime drops to $\sim$10\,s.  For comparison, CAMB
at its default accuracy (\texttt{AccuracyBoost=1}) takes under a second for
the same cosmology.  The speed difference reflects the inherent overhead of
a pure Python implementation relative to compiled Fortran;
\texttt{nanoCMB} is not intended to compete on speed, but rather to
provide a readable, self-contained reference implementation.
The first run incurs an additional one-off cost to build the spherical
Bessel function lookup tables, which are then cached to disk and reused
in subsequent runs.

\section{Discussion and conclusions}
\label{sec:conclusions}

We have presented \texttt{nanoCMB}, a pedagogical CMB power spectrum
calculator that achieves sub-percent accuracy relative to CAMB in
approximately 1400 lines of Python.  The code is designed to be read
alongside this paper, with each section corresponding to a clearly
delineated block of the source code.

\paragraph{Design philosophy.}
Several deliberate choices keep the code readable:
\begin{itemize}
\item Single file: the entire calculation lives in \texttt{nanocmb.py},
with no package structure or build system.
\item Dictionary-based state: background and thermodynamic quantities are
passed as plain Python dictionaries, avoiding class hierarchies.
\item Minimal dependencies: the only requirements are NumPy and SciPy
(with optional Numba for acceleration).
\item Explicit over implicit: every numerical choice (grid sizes,
tolerances, truncation orders) is a visible constant or parameter, not hidden
in a configuration file.
\end{itemize}

\paragraph{Limitations.}
The current implementation omits several effects that are important for
precision cosmology:
\begin{itemize}
\item Gravitational lensing of the CMB by large-scale structure, which
smooths the acoustic peaks and generates B-mode polarisation.
\item Tensor perturbations (gravitational waves), which contribute to
B-modes at large scales.
\item Massive neutrinos, which affect the matter power spectrum and
CMB lensing.
\item Spatial curvature ($K \neq 0$).
\item Non-standard dark energy ($w \neq -1$).
\end{itemize}

\paragraph{Possible extensions.}
The modular structure makes extensions relatively straightforward.
Gravitational lensing would require computing the lensing potential power
spectrum $C_\ell^{\phi\phi}$ and applying the lensing convolution.
Tensor perturbations require a separate Boltzmann hierarchy for
gravitational wave modes.  Massive neutrinos require momentum-dependent
distribution functions.  Each of these extensions would add on the order
of 100--200 lines, reusing the existing infrastructure (grids, Bessel
tables, ODE solver).

\appendix

\section{Atomic data}
\label{app:atomic}

Table~\ref{tab:atomic} lists the atomic constants used in the recombination
calculation.

\begin{table}[tbp]
\centering
\begin{tabular}{llc}
\toprule
Quantity & Symbol & Value \\
\midrule
H ionisation & $L_{\rm H,ion}$ & $1.096788 \times 10^7\,$m$^{-1}$ \\
H Lyman-$\alpha$ & $L_{\rm H,\alpha}$ & $8.225916 \times 10^6\,$m$^{-1}$ \\
He\,{\sc i} ionisation & $L_{\rm He1,ion}$ & $1.983108 \times 10^7\,$m$^{-1}$ \\
He\,{\sc ii} ionisation & $L_{\rm He2,ion}$ & $4.389089 \times 10^7\,$m$^{-1}$ \\
He\,{\sc i} $2^1S_0$ & $L_{\rm He,2s}$ & $1.662774 \times 10^7\,$m$^{-1}$ \\
He\,{\sc i} $2^1P_1$ & $L_{\rm He,2p}$ & $1.711349 \times 10^7\,$m$^{-1}$ \\
H two-photon decay & $\Lambda_{2s \to 1s}$ & 8.2246\,s$^{-1}$ \\
He\,{\sc i} two-photon decay & $\Lambda_{\rm He}$ & 51.3\,s$^{-1}$ \\
He\,{\sc i} $2^1P_1$ decay & $A_{2P,s}$ & $1.798 \times 10^9\,$s$^{-1}$ \\
He\,{\sc i} $2^3P_1$ decay & $A_{2P,t}$ & 177.58\,s$^{-1}$ \\
Thomson cross section & $\sigma_T$ & $6.652 \times 10^{-29}\,$m$^2$ \\
\bottomrule
\end{tabular}
\caption{\label{tab:atomic}
Atomic transition data used in the RECFAST recombination calculation.}
\end{table}

\section{State vector layout}
\label{app:statevec}

The perturbation state vector $\mathbf{y}$ is a flat array of $N_{\rm var} = 50$
elements, laid out as follows:

\begin{table}[htbp]
\centering
\footnotesize
\begin{tabular}{clll}
\toprule
Index & Variable & Symbol & Description \\
\midrule
0 & \texttt{etak} & $k\eta$ & Metric \\
1 & \texttt{clxc} & $\delta_c$ & CDM density \\
2 & \texttt{clxb} & $\delta_b$ & Baryon density \\
3 & \texttt{vb} & $v_b$ & Baryon velocity \\
4--19 & \texttt{G[0..15]} & $\Theta_0\!\ldots\!\Theta_{15}$ & Photon temp. \\
20--33 & \texttt{POL[0..13]} & $E_2\!\ldots\!E_{15}$ & Photon pol. \\
34--49 & \texttt{R[0..15]} & $\mathcal{N}_0\!\ldots\!\mathcal{N}_{15}$ & Neutrino \\
\bottomrule
\end{tabular}
\caption{\label{tab:statevec}
State vector layout for the Boltzmann hierarchy ODE system.  Total: 50 variables.}
\end{table}

\section*{Pipeline overview}
\label{sec:pipeline}

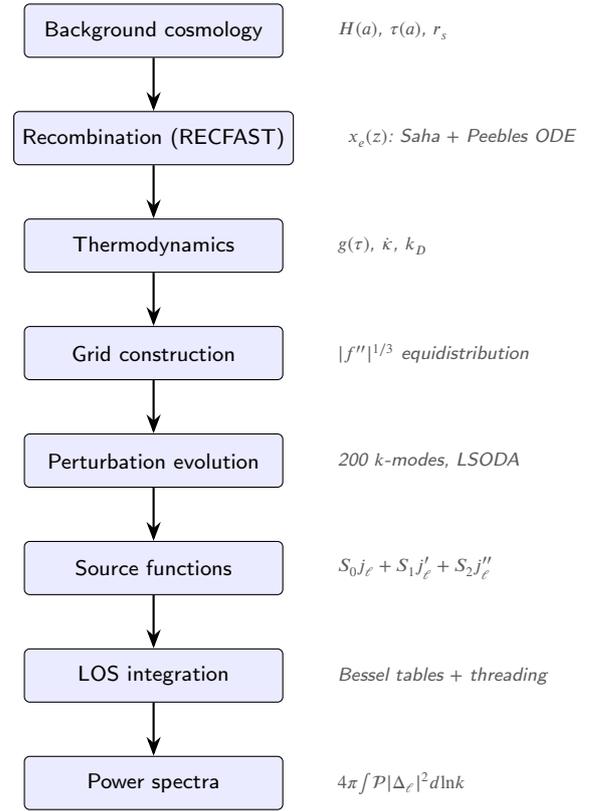
\begin{figure}[!htbp]
\centering
\begin{tikzpicture}[
  node distance=0.7cm and 0.6cm,
  block/.style={rectangle, draw, fill=blue!8, minimum width=3.4cm, minimum height=0.7cm,
                align=center, font=\footnotesize, rounded corners=2pt},
  arrow/.style={-{Stealth[length=2.5mm]}, thick},
  label/.style={font=\scriptsize\itshape, text=gray!70!black},
]
\node[block] (bg) {Background cosmology};
\node[block, below=of bg] (recomb) {Recombination (RECFAST)};
\node[block, below=of recomb] (thermo) {Thermodynamics};
\node[block, below=of thermo] (grids) {Grid construction};
\node[block, below=of grids] (pert) {Perturbation evolution};
\node[block, below=of pert] (src) {Source functions};
\node[block, below=of src] (los) {LOS integration};
\node[block, below=of los] (cls) {Power spectra};

\draw[arrow] (bg) -- (recomb);
\draw[arrow] (recomb) -- (thermo);
\draw[arrow] (thermo) -- (grids);
\draw[arrow] (grids) -- (pert);
\draw[arrow] (pert) -- (src);
\draw[arrow] (src) -- (los);
\draw[arrow] (los) -- (cls);

\node[label, right=0.6cm of bg] {$H(a)$, $\tau(a)$, $r_s$};
\node[label, right=0.6cm of recomb] {$x_e(z)$: Saha $+$ Peebles ODE};
\node[label, right=0.6cm of thermo] {$g(\tau)$, $\dot\kappa$, $k_D$};
\node[label, right=0.6cm of grids] {$|f''|^{1/3}$ equidistribution};
\node[label, right=0.6cm of pert] {200 $k$-modes, LSODA};
\node[label, right=0.6cm of src] {$S_0 j_\ell + S_1 j_\ell' + S_2 j_\ell''$};
\node[label, right=0.6cm of los] {Bessel tables $+$ threading};
\node[label, right=0.6cm of cls] {$4\pi\!\int\! \mathcal{P}|\Delta_\ell|^2 d\!\ln\!k$};

\end{tikzpicture}
\caption{\label{fig:pipeline}
Schematic of the \texttt{nanoCMB} computational pipeline, from cosmological
parameters to angular power spectra.}
\end{figure}

Figure~\ref{fig:pipeline} shows a schematic overview of the \texttt{nanoCMB}
computational pipeline.

\section*{Acknowledgements}

We thank Antony Lewis for developing and maintaining CAMB, which made the detailed validation in this work possible.
The work of AM was supported by an STFC Consolidated Grant [Grant No.\ ST/X000672/1].
This work made use of CAMB~\cite{Lewis:1999bs} for validation, and
NumPy~\cite{Harris:2020xlr} and SciPy~\cite{Virtanen:2019joe} for numerical computation.

\section*{Declaration of competing interest}

The author declares that there are no known competing financial interests or personal relationships that could have appeared to influence the work reported in this paper.

\section*{Data availability}

The \texttt{nanoCMB} code, validation scripts, and all data needed to reproduce the figures in this paper are available at \url{https://github.com/adammoss/nanoCMB}.

\printcredits

\bibliographystyle{cas-model2-names}
\bibliography{refs}

@article{Planck:2018vyg,
    author = "Aghanim, N. and others",
    collaboration = "Planck",
    title = "{Planck 2018 results. VI. Cosmological parameters}",
    eprint = "1807.06209",
    archivePrefix = "arXiv",
    primaryClass = "astro-ph.CO",
    doi = "10.1051/0004-6361/201833910",
    journal = "Astron. Astrophys.",
    volume = "641",
    pages = "A6",
    year = "2020",
    note = "[Erratum: Astron.Astrophys. 652, C4 (2021)]"
}

@article{Lewis:1999bs,
    author = "Lewis, Antony and Challinor, Anthony and Lasenby, Anthony",
    title = "{Efficient computation of CMB anisotropies in closed FRW models}",
    eprint = "astro-ph/9911177",
    archivePrefix = "arXiv",
    doi = "10.1086/309179",
    journal = "Astrophys. J.",
    volume = "538",
    pages = "473--476",
    year = "2000"
}

@article{Blas:2011rf,
    author = "Blas, Diego and Lesgourgues, Julien and Tram, Thomas",
    title = "{The Cosmic Linear Anisotropy Solving System (CLASS) II: Approximation schemes}",
    eprint = "1104.2933",
    archivePrefix = "arXiv",
    primaryClass = "astro-ph.CO",
    reportNumber = "CERN-PH-TH-2011-082, LAPTH-010-11",
    doi = "10.1088/1475-7516/2011/07/034",
    journal = "JCAP",
    volume = "07",
    pages = "034",
    year = "2011"
}

@article{Seljak:1996is,
    author = "Seljak, Uros and Zaldarriaga, Matias",
    title = "{A Line of sight integration approach to cosmic microwave background anisotropies}",
    eprint = "astro-ph/9603033",
    archivePrefix = "arXiv",
    doi = "10.1086/177793",
    journal = "Astrophys. J.",
    volume = "469",
    pages = "437--444",
    year = "1996"
}

@article{Ma:1995ey,
    author = "Ma, Chung-Pei and Bertschinger, Edmund",
    title = "{Cosmological perturbation theory in the synchronous and conformal Newtonian gauges}",
    eprint = "astro-ph/9506072",
    archivePrefix = "arXiv",
    doi = "10.1086/176550",
    journal = "Astrophys. J.",
    volume = "455",
    pages = "7--25",
    year = "1995"
}

@article{Bertschinger:1995er,
    author = "Bertschinger, Edmund",
    title = "{COSMICS: cosmological initial conditions and microwave anisotropy codes}",
    eprint = "astro-ph/9506070",
    archivePrefix = "arXiv",
    reportNumber = "GC3-033",
    month = "6",
    year = "1995"
}

@article{Seager:1999bc,
    author = "Seager, Sara and Sasselov, Dimitar D. and Scott, Douglas",
    title = "{A new calculation of the recombination epoch}",
    eprint = "astro-ph/9909275",
    archivePrefix = "arXiv",
    doi = "10.1086/312250",
    journal = "Astrophys. J. Lett.",
    volume = "523",
    pages = "L1--L5",
    year = "1999"
}

@article{Seager:1999km,
    author = "Seager, Sara and Sasselov, Dimitar D. and Scott, Douglas",
    title = "{How exactly did the universe become neutral?}",
    eprint = "astro-ph/9912182",
    archivePrefix = "arXiv",
    reportNumber = "UBC-COS-99-07",
    doi = "10.1086/313388",
    journal = "Astrophys. J. Suppl.",
    volume = "128",
    pages = "407--430",
    year = "2000"
}

@article{Pequignot1991,
    author = "Pequignot, D. and Petitjean, P. and Boisson, C.",
    title = "{Total and effective radiative recombination coefficients}",
    doi = "10.1051/0004-6361:19910127",
    journal = "Astron. Astrophys.",
    volume = "251",
    pages = "680--688",
    year = "1991"
}

@article{HummerStorey1998,
    author = "Hummer, D. G. and Storey, P. J.",
    title = "{Recombination of helium-like ions --- I. Photoionization cross-sections and total recombination and cooling coefficients for atomic helium}",
    doi = "10.1046/j.1365-8711.1998.01596.x",
    journal = "Mon. Not. Roy. Astron. Soc.",
    volume = "297",
    pages = "1073--1078",
    year = "1998"
}

@article{Scott:2009sz,
    author = "Scott, Douglas and Moss, Adam",
    title = "{Matter temperature after cosmological recombination}",
    eprint = "0902.3438",
    archivePrefix = "arXiv",
    primaryClass = "astro-ph.CO",
    doi = "10.1111/j.1365-2966.2009.14939.x",
    journal = "Mon. Not. Roy. Astron. Soc.",
    volume = "397",
    pages = "445--446",
    year = "2009"
}

@article{Wong:2007ym,
    author = "Wong, Wan Yan and Moss, Adam and Scott, Douglas",
    title = "{How well do we understand cosmological recombination?}",
    eprint = "0711.1357",
    archivePrefix = "arXiv",
    primaryClass = "astro-ph",
    doi = "10.1111/j.1365-2966.2008.13092.x",
    journal = "Mon. Not. Roy. Astron. Soc.",
    volume = "386",
    pages = "1023--1028",
    year = "2008"
}

@article{Peebles:1968ja,
    author = "Peebles, P. J. E.",
    title = "{Recombination of the Primeval Plasma}",
    doi = "10.1086/149628",
    journal = "Astrophys. J.",
    volume = "153",
    pages = "1",
    year = "1968"
}

@article{Zeldovich1968,
    author = "Zel'dovich, Ya. B. and Kurt, V. G. and Syunyaev, R. A.",
    title = "{Recombination of Hydrogen in the Hot Model of the Universe}",
    journal = "Zh. Eksp. Teor. Fiz.",
    volume = "55",
    pages = "278",
    year = "1968",
    note = "[Sov. Phys. JETP {\bf 28} (1969) 146]"
}

@book{Dodelson:2020bqr,
    author = "Dodelson, Scott and Schmidt, Fabian",
    title = "{Modern Cosmology}",
    doi = "10.1016/C2017-0-01943-2",
    publisher = "Academic Press",
    year = "2020"
}

@article{Callin:2006qx,
    author = "Callin, Petter",
    title = "{How to calculate the CMB spectrum}",
    eprint = "astro-ph/0606683",
    archivePrefix = "arXiv",
    month = "6",
    year = "2006"
}

@article{Akima1970,
    author = "Akima, Hiroshi",
    title = "{A New Method of Interpolation and Smooth Curve Fitting Based on Local Procedures}",
    doi = "10.1145/321607.321609",
    journal = "J. ACM",
    volume = "17",
    number = "4",
    pages = "589--602",
    year = "1970"
}

@article{Harris:2020xlr,
    author = "Harris, Charles R. and others",
    title = "{Array programming with NumPy}",
    eprint = "2006.10256",
    archivePrefix = "arXiv",
    primaryClass = "cs.MS",
    doi = "10.1038/s41586-020-2649-2",
    journal = "Nature",
    volume = "585",
    number = "7825",
    pages = "357--362",
    year = "2020"
}

@article{Virtanen:2019joe,
    author = "Virtanen, Pauli and others",
    title = "{SciPy 1.0--Fundamental Algorithms for Scientific Computing in Python}",
    eprint = "1907.10121",
    archivePrefix = "arXiv",
    primaryClass = "cs.MS",
    doi = "10.1038/s41592-019-0686-2",
    journal = "Nature Meth.",
    volume = "17",
    pages = "261",
    year = "2020"
}

\end{document}